\documentstyle[12pt,aasms4,psfig]{article}
\def\ltsima{$\; \buildrel < \over \sim \;$}
\def\simlt{\lower.5ex\hbox{\ltsima}}
\def\gtsima{$\; \buildrel > \over \sim \;$}
\def\simgt{\lower.5ex\hbox{\gtsima}}
\def\s{\ifmmode \widetilde \else \~\fi}
\def\={\overline}

\def\spose#1{\hbox to 0pt{#1\hss}}

\def\lta{\mathrel{\spose{\lower 3pt\hbox{$\mathchar"218$}}
     \raise 2.0pt\hbox{$\mathchar"13C$}}}
\def\gta{\mathrel{\spose{\lower 3pt\hbox{$\mathchar"218$}}
     \raise 2.0pt\hbox{$\mathchar"13E$}}}
\def\Dt{\spose{\raise 1.5ex\hbox{\hskip3pt$\mathchar"201$}}}	
\def\dt{\spose{\raise 1.0ex\hbox{\hskip2pt$\mathchar"201$}}}	

\def\=={\equiv}

\def\dotsfill{\leaders\hbox to 1em{\hss.\hss}\hfill}

\def\Gyr{{\rm\,Gyr}}

\received{4 July 2001}

\lefthead{Richer}
\righthead{}

\begin{document}

\title{SEARCHING FOR THE GALACTIC DARK MATTER}

\author{
 Harvey B. Richer}

\vspace{1in}
 
\centerline {{Department of Physics \& Astronomy, University of British Columbia}}

\centerline {6224 Agricultural Road, Vancouver, B.C., V6T 1Z1, Canada}
\centerline {Electronic mail: richer@astro.ubc.ca}

\vspace{1in}

\centerline{To appear in proceedings of ``The Dark Universe: Matter, Energy and Gravity"}

\centerline{held at the Space Telescope Science Institute April 2 $-$ 5 2001.}



\begin{abstract}
A straightforward interpretation of the MACHO microlensing results in the
direction of the Magellanic Clouds suggests that an important fraction
of the baryonic dark matter component of our Galaxy is in the form of old
white dwarfs. If correct, this has serious implications for the early
generations of stars that formed in the Universe and also on the manner in which
galaxies formed and enriched themselves in heavy elements. I examine this scenario in some 
detail and in particular explore
whether the searches currently being carried out to locate local examples
of these MACHOs can shed any light at all on this scenario.

\end{abstract}

\section{Introduction}

A conservative estimate of the mass of the Galaxy out to a distance of about 2/3
of that of the Large Magellanic Cloud is $\rm M_G = 4 \times 10^{11} M_{\odot}$ (Fich and Tremaine 1991). With a total luminosity in the V-band of $\rm 1.4 \times 10^{10} L_{\odot}$ (Binney and Tremaine 1987) the Galactic mass to light ratio in V ($ \rm M/L_V$) out to $\rm 35$ kpc is $\rm \sim 30$. Since normal
stellar populations do not generally produce $\rm M/L_V$ ratios higher than about 3, this is
usually taken as evidence for an important component of dark matter within an extended halo surrounding the Galaxy.

A related result is that of the MACHO microlensing experiment in the direction of the Magellanic Clouds which seems to indicate that about 20\% ($10 - 50$\%
is the 90\% CI) of the dark matter in
the Galaxy is tied up in objects with masses near $\rm 0.5~M_{\odot}$ (0.3 - 0.9 is the 90\% CI) (Alcock 2000).
This assertion is only true if the bulk of the MACHOs are located in the halo of our Galaxy. Moreover, recent studies suggest that objects with masses from $\rm 10^{-7}$ to $\rm 0.02~M_{\odot}$ are now largely excluded (so planets and/or brown dwarfs are not important dark matter candidates) and that a 100\% MACHO halo is no longer a viable model (Alcock 2000). 

MACHO masses near $\rm 0.5~M_{\odot}$ are suggestive of white dwarfs although neutron stars or even primordial black holes remain as unlikely but still possible candidates.
A halo currently consisting of 20\% old white dwarfs implies that the precursors of these $\rm 0.5~M_{\odot}$ objects would have accounted for $\rm \sim 40\%$  
of $\rm \Omega_{\rm baryon}$ in the Universe. 
This estimate comes from taking $\rm 2~M_{\odot}$ as the precursor masses, $\rm M/L_V = 1500$ for $\rm \Omega_{critical}$, 0.04 for $\rm \Omega_{baryon}$ and
assuming that the Galactic halo is a fair sample of the types of mass seen in the Universe. It may be that Galactic halos are baryon-rich in which case the derivation of 40\% would be an over-estimate. This may also
be an interesting way to constrain the masses of the precursors as clearly in this analysis they could not exceed $\rm \sim 5~M_{\odot}$ without violating the observed baryonic content of the Universe. Even given all the uncertainties in these estimates, it is clear in this scenario
that an important fraction of all the baryons in the Universe were funnelled through
this star formation mode. This will surely have implications for star formation
scenarios in the early Universe and perhaps also on the manner in which galaxies
were assembled and how they enriched themselves in heavy elements. 
 
The interpretation of the microlensing results is, however,  not without controversy as there are indications that at least some of the lenses may reside
in the Magellanic Clouds themselves (McGrath and Sahu 2000; Sahu 2002). 
This would have the effect of lowering
both the lens masses (to $\rm 0.2 - 0.3~M_{\odot}$) and the contribution of the
MACHOs to the Galactic dark matter budget. The lower lens masses would be in
line with expectations from mass function considerations and the lower MACHO
dark matter contribution would alleviate such problems as the appearance of
high redshift galaxies (at an epoch when the white dwarf precursors were luminous), the chemical evolution of galaxies (the white dwarf precursors
producing too much helium and heavier elements) and the peculiar mass
function required of the precursors (a mass function truncated at both high
and low masses to avoid too much chemical enrichment from high mass stars
and too many halo white dwarfs currently evolving from the low mass ones [Chabrier 1999]).
There is also the possibility that most of the MACHOs are located in a
thick (Gates {\it et al.} 1998)
or flaring disk (Evans {\it et al.} 1998) of the Galaxy. This would
have the effect of reducing their scale height and again their total Galactic
mass contribution.

The definitive answer as to whether old white dwarfs are important baryonic
dark matter contributors will come from searches which will (or will not) identify significant numbers of high velocity local examples of the very cool white dwarfs that produce the microlensing. Such searches are now underway
and in the ensuing sections I will discuss early results from them.

\section{New Models of Very Cool White Dwarfs}

Just recently, a new era opened in the study of old white dwarfs. Hansen
(1998, 1999) first 
calculated emergent spectra from cool ($\rm T_{eff} < 4000$K)
white dwarfs with atmospheres that included opacity from the $\rm H_2$ molecule. Similar models were also constructed by Saumon and Jacobson (1999) and 
earlier ones by Bergeron, Saumon and Wesemael (1995) for objects down to
4000K. All these
models exhibited very strong collisionally induced opacity in the red and near IR part of the
spectrum and the effects of this opacity can be seen clearly in the model spectrum shown in Figure 1
as well in several dozen real objects (see e.g. Oppenheimer {\it et al}. 2001a).
The broad bands seen in the spectrum in Figure 1 are due to $\rm H_2$ but it is of interest
to point out that no cool white dwarf has yet been observed that actually
shows these spectral features. Perhaps it is because no objects this cool have
been discovered up until now, or that the models are still somehow
deficient, or that possibly the Universe is not old enough for
hydrogen-rich white dwarfs to have reached such low temperatures. The effects
of collisionally induced absorption are not limited, however, to extremely
cool objects. Its influence can been seen in white dwarfs at least as hot as
5400K (see, e.g., the spectrum of LHS 1126 in Bergeron {\it et al.} 1994).

\begin{figure}
\psfig{figure=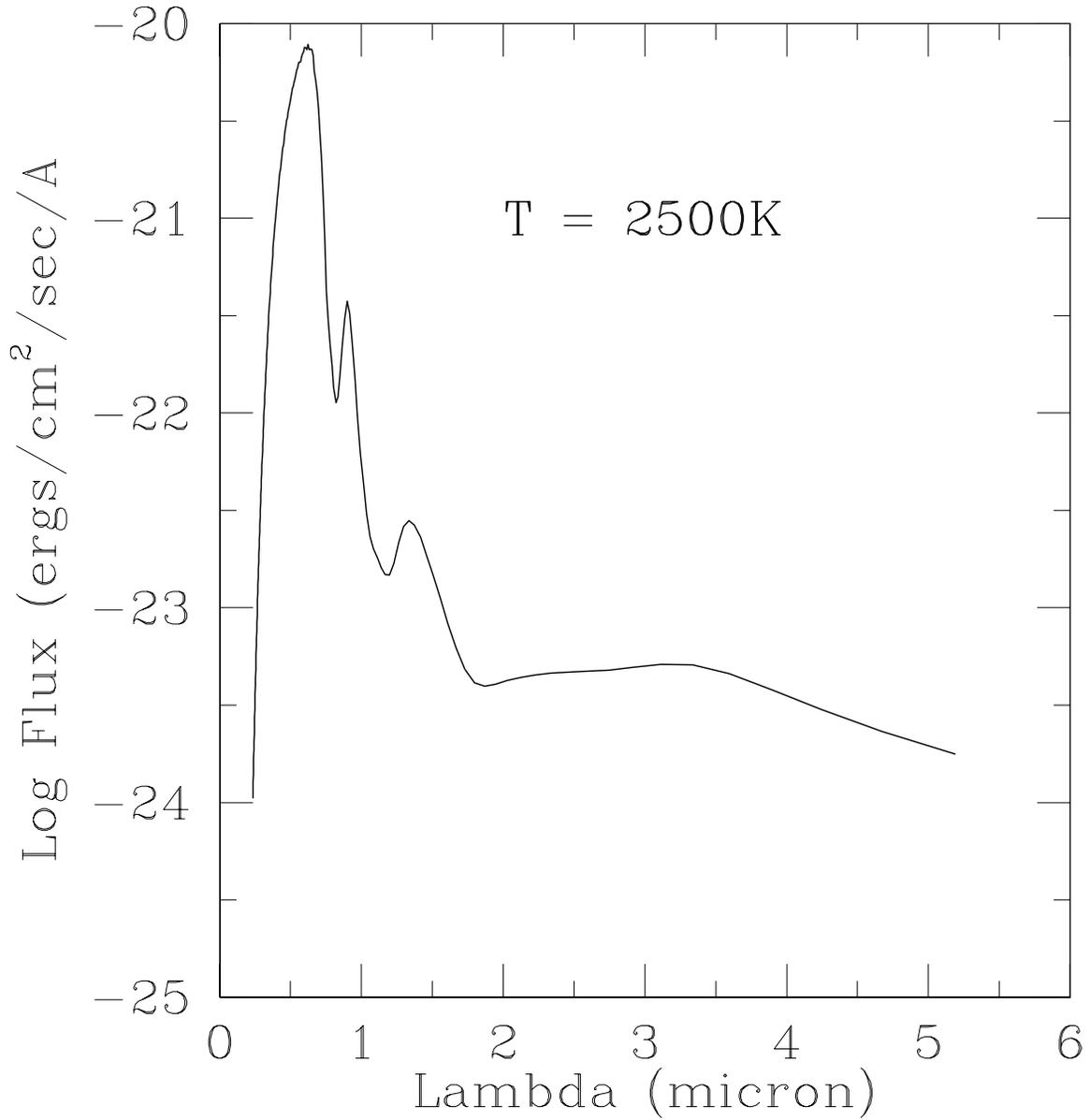,width=\hsize}
\caption{The spectrum of a 2500K hydrogen-rich white dwarf from Hansen (1999, private communication). The depletion of flux in the near IR has been seen now in a dozen or more old white
dwarfs. No known object, however, exhibits significant $\rm H_2$ bands
as illustrated in this model spectrum.}
\end{figure}

The effect of the $\rm H_2$ opacity is to force the radiation out in the  bluer
spectral regions. This has an enormous influence on their colors, 
the location of the white dwarf cooling track in the color-magnitude diagram,
and hence their observability.
Down to a temperature of about 4000K (which corresponds to
an age of about $7 \Gyr$ for $\rm 0.5~M_{\odot}$ white dwarfs) these  stars become increasingly redder as they cool, the reddest color they achieve is about
${\rm V   - I \simeq  1.2}$ in Hansen's (1998, 1999) models. Note that other sets of cooling models give somewhat different colors and ages here (e.g. Fontaine {\it et al.} 2001).
Older white dwarfs
become progessively bluer in $\rm V - I$ as they continue to cool.
Ancient white dwarfs,  of age $\rm 12 \Gyr$,  and
mass $\rm 0.5~M_{\odot}$,   have  ${\rm  V  -   I  \sim  -0.3}$ according  to   these
models. As has been noted several times in the past few years, {\it old hydrogen-rich white dwarfs are blue, not red, in ${\it V-I}$ colors.} Searches for such ancient
objects are now concentrating on bluish objects in these colors. 

A cooling track from Hansen's (1998, 1999) models illustrating all these features for $\rm 0.5~M_{\odot}$
white dwarfs is shown in Figure 2. Note that instead of getting arbitrarily
faint with age, old hydrogen-rich white dwarfs seem eventually to coast at approximately constant
$\rm M_V$ for ages between about 10 and 14 Gyr, just the timescales
which may
bracket globular cluster ages. The implication is clear $-$ very old white
dwarfs may be easier to detect than was previously thought. Also potentially of interest is that
the $\rm V - I$ color now  
moves the bulk of the ancient white dwarfs away from the
swarm of field K and M-dwarfs (making confusion with such objects $\rm {\it less}$ likely)
but at the same time overlapping the colors of distant star-forming galaxies
(making the confusion with such objects $\rm {\it more}$ likely). The solution
here will be to search for old white dwarfs using both color and proper motion
criteria, an approach taken by most current programs.

\begin{figure}
\psfig{figure=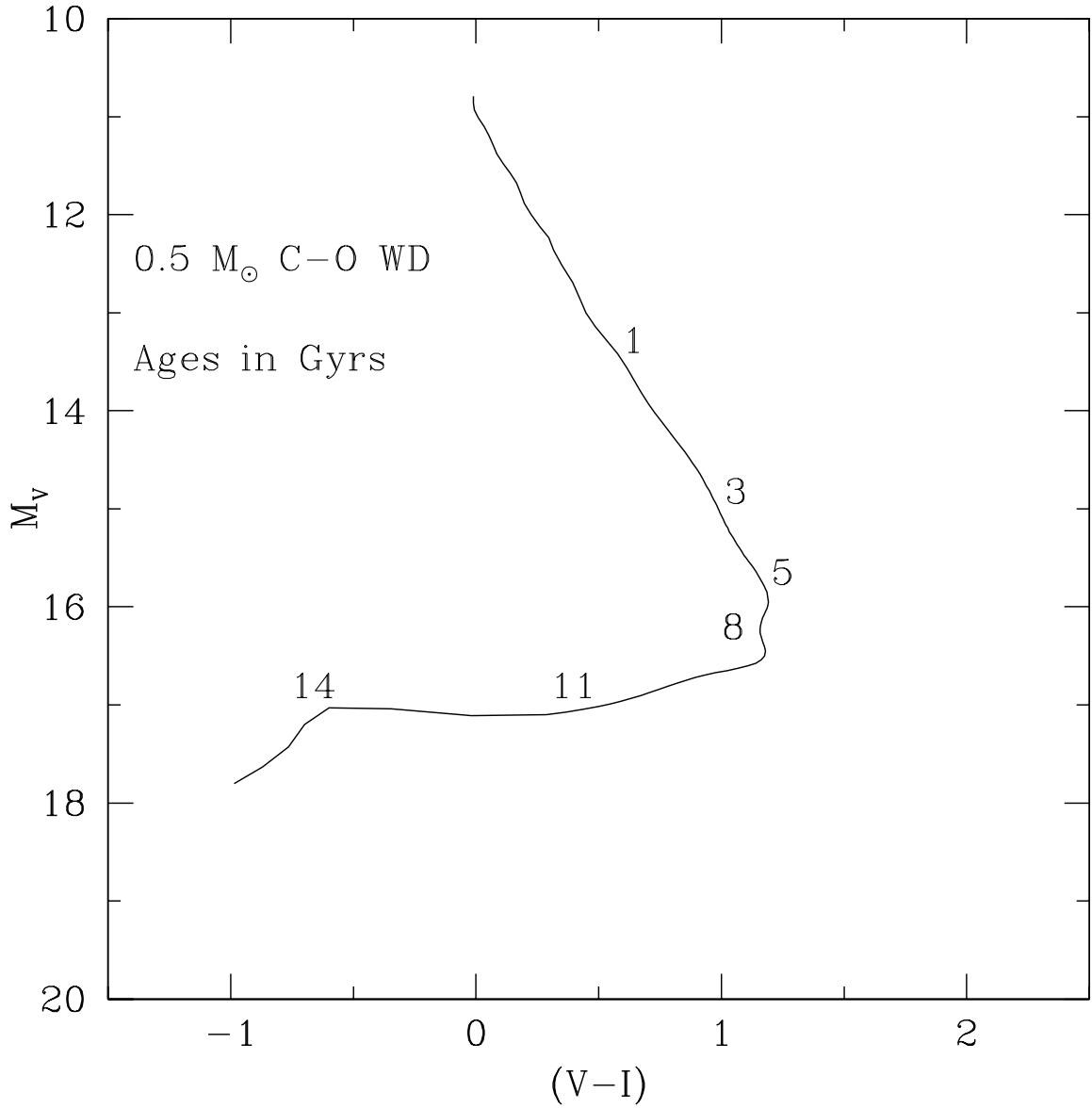,width=\hsize}
\caption{Cooling track of a $\rm 0.5~M_{\odot}$ white dwarf (Hansen 1999). The age of the white dwarf in Gyrs
is indicated along the plot. This age does not include the main sequence lifetime of the white dwarf precursor.}
\end{figure}

One problem with these models is that they are largely untested. Of serious
concern is that none of the cool white dwarfs found to date (see following sections) seem to exhibit the extreme blue colors predicted by most of
these current models containing $\rm H_2$ in their atmospheres. This may be
suggesting missing or inadequate physics. One place where these models could
potentially be tested is among the cool white dwarfs in globular clusters.
Such systems should contain white dwarfs almost as old as the clusters
themselves (likely in the $\rm 12 - 14$ Gyr range), so the extreme effects of
collisionally induced opacity ought to be clearly seen. Figure 3 shows the
current situation here. It displays the deepest color magnitude diagram
yet obtained for a globular cluster (Richer {\it et al.} 1995, 1997), an HST
study of M4. White dwarfs perhaps as faint as $\rm M_V = 16$ are recorded in this study. If these  objects have masses near $\rm 0.5~M_{\odot}$, they
will have ages in excess of 7 Gyr and $\rm T_{eff} \sim 4000K$ in Hansen's (1998, 1999) models; cool enough to exhibit some of the effects of collisionally induced opacity but not old enough to help constrain the appearance of truly ancient white dwarfs. In HST cycle 9, a group of colleagues and I
were awarded time on HST to attempt to locate the termination point of the white
dwarf cooling sequence in M4. With the current generation of models and a
cluster age in the range of $\rm 12 - 14$ Gyr, this should occur near a V magnitude
of about 30, certainly a enormous challenge. It is not clear whether the
data will actually go this deep, but even if it does not, we should be
in a position to provide some stringent tests of the current generation of models.

\begin{figure}
\psfig{figure=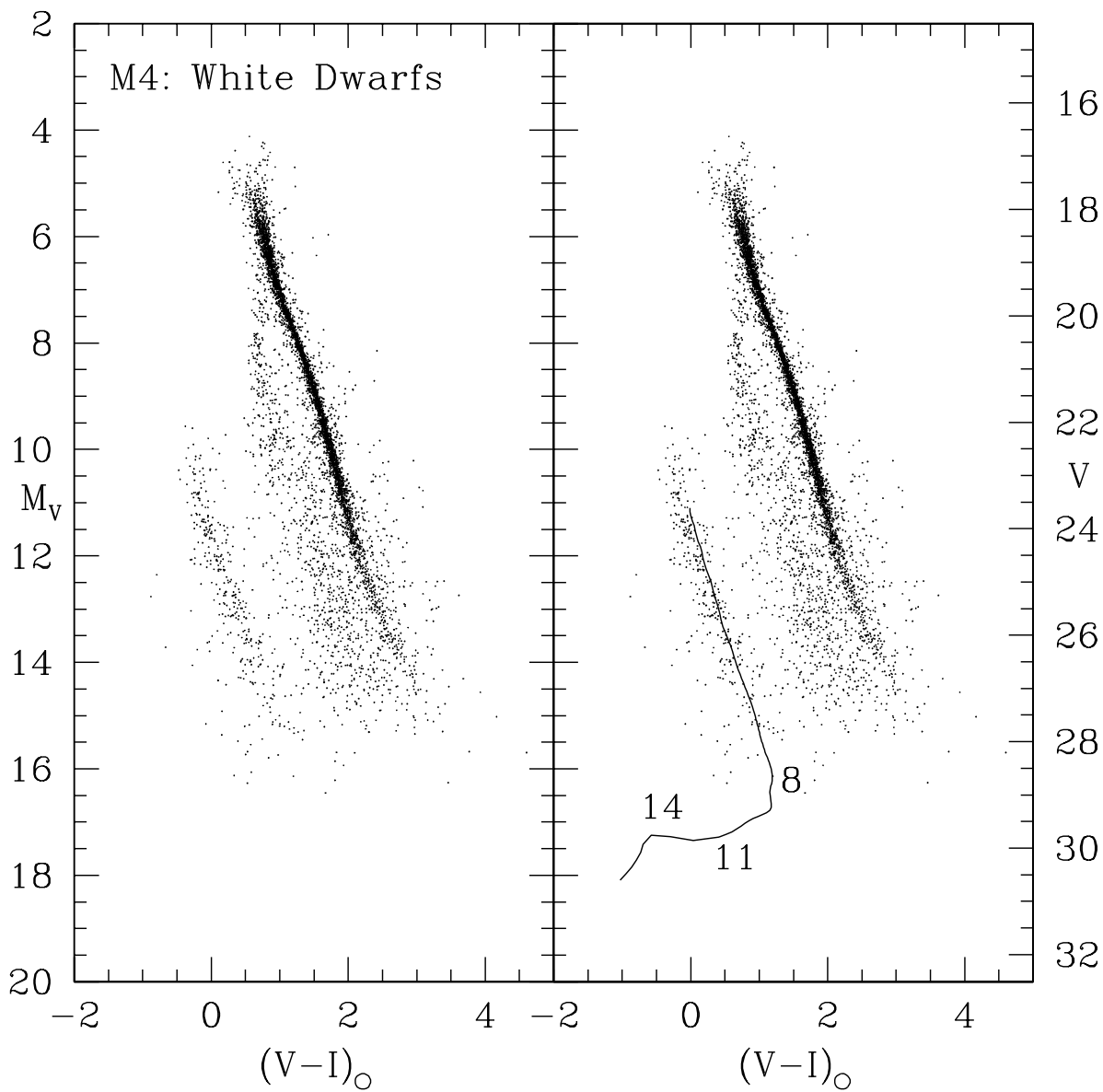,width=\hsize}
\caption{Deep HST color magnitude diagram of the globular cluster M4 with
a cooling track for a $\rm 0.5~M_{\odot}$ white dwarf superimposed. The oldest
white dwarfs detected are only about 7 Gyr and while these should show some effects due to collisionally induced absorption, they are too warm to exhibit the extreme effect of a bluing in the $\rm V - I$ color with decreasing temperature.}
\end{figure}

In Figure 4 I attach a simulation of what we expect to see with these data.
This is an optimistic simulation in that it does not include the effects of
charge transfer inefficiency now known to be plaguing the HST WFPC2
CCDs and also assumes that all the data are taken under low sky brightness
conditions. Nevertheless, even if the faintest magnitudes are not reached,
the effects of collisionally induced opacity should be readily apparent and quantitatively testable with these data.

 \begin{figure}[h]
\plottwo{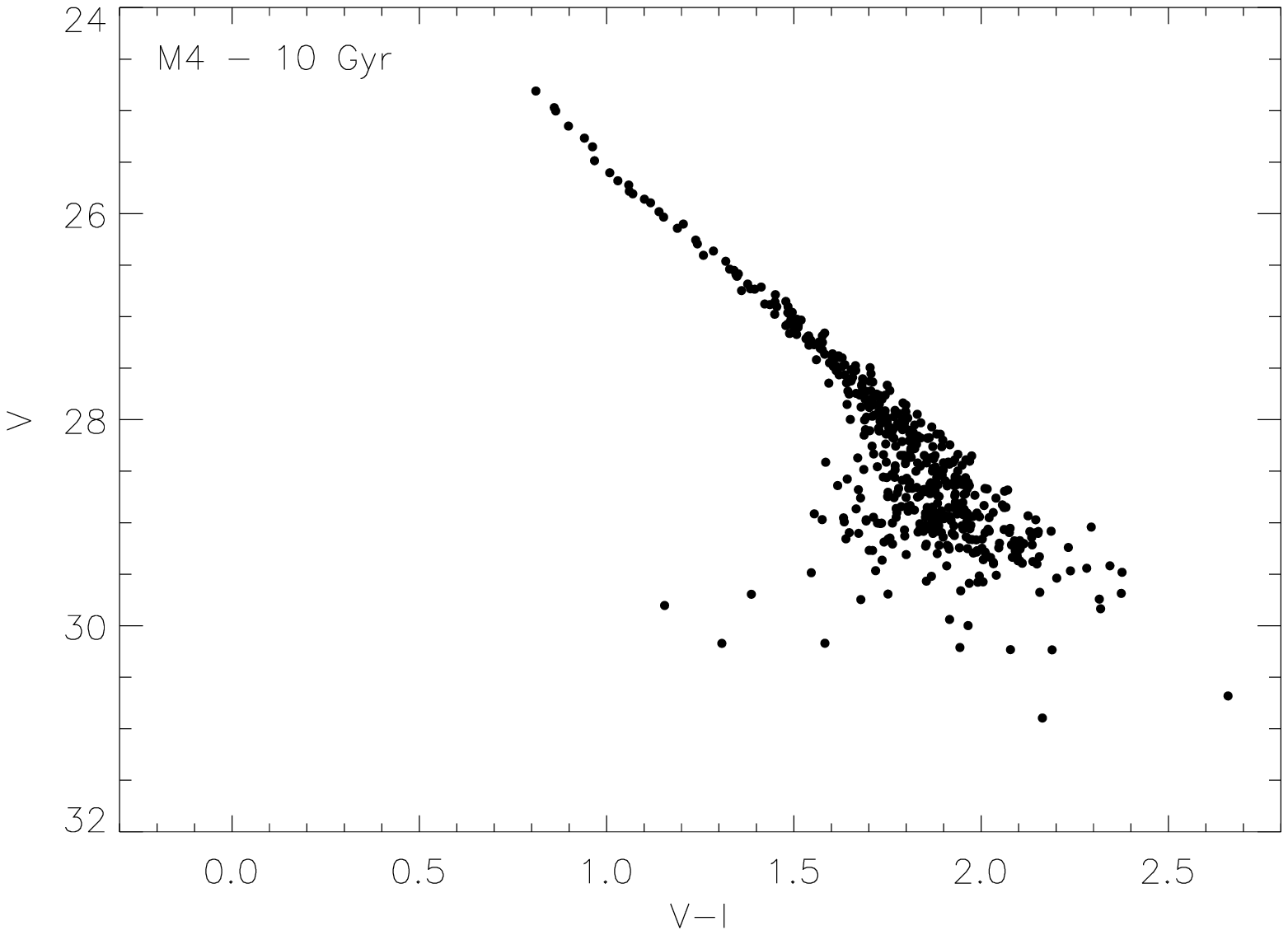}{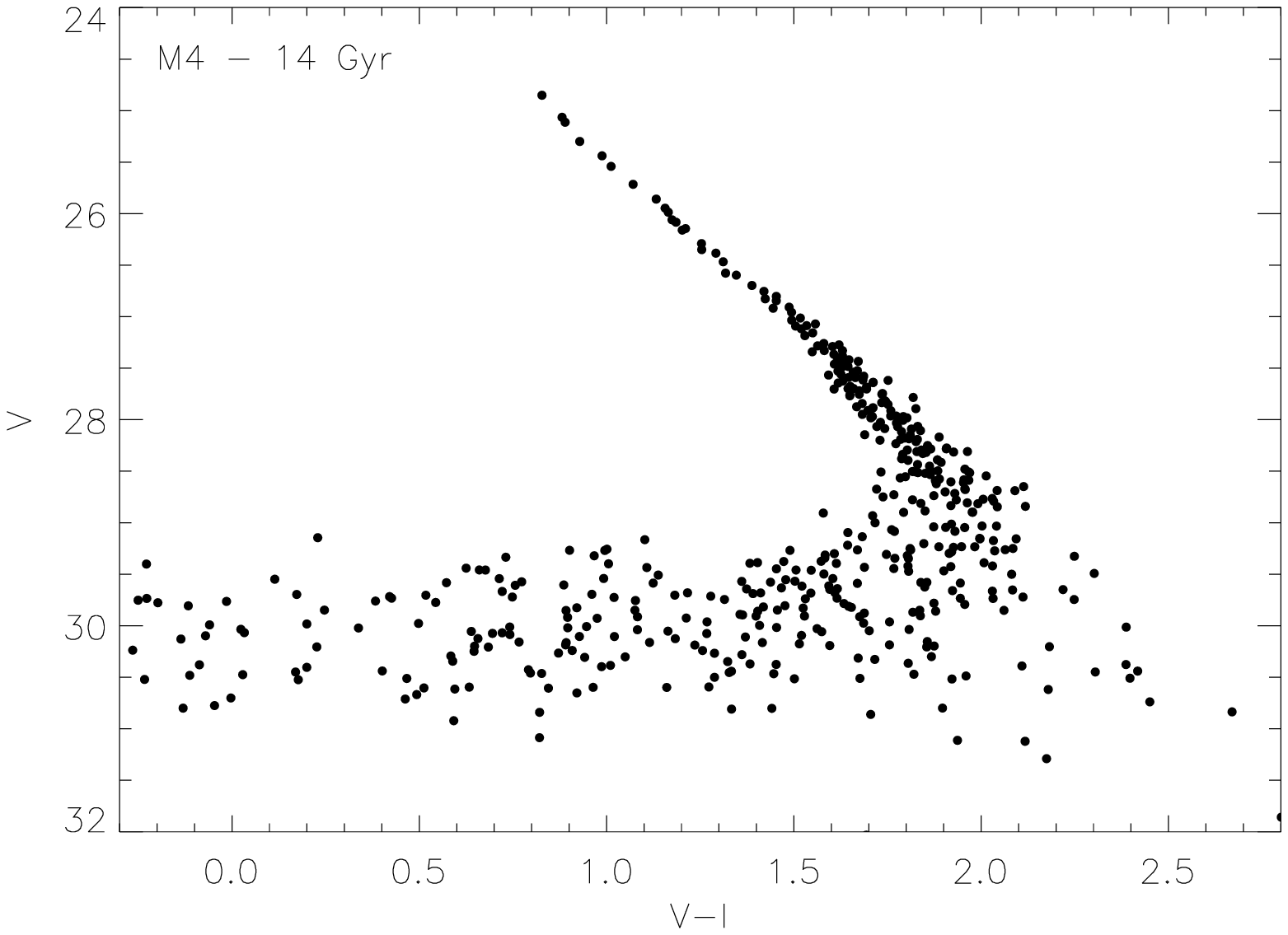}
\caption{Simulation of deep HST data in the globular cluster M4.
The WFPC2 exposure time calculator was used to determine the photometric uncertainty as a function of apparent magnitude for the white dwarfs in the globular cluster M4 assuming ages of 10 and 14 Gyr. A suite of 
Monte Carlo realizations, incorporating a proper treatment of the photometric
uncertainties involved and the higher background expected from scattered
light of the bright giants, were then undertaken. One such realization
 is shown for a cluster with each of these ages.}
\label{twobarrel}
\end{figure}

\section{Current Searches for Very Cool White Dwarfs}

A number of searches for an ancient population of halo white dwarfs
are currently underway or have been completed. Most of these involve color
and proper motion selection with some follow-up spectroscopy. The proper
motions have generally been derived from digitized photographic plates with
quite long time baselines (typically $\rm 20 - 50$ years). I list below in Table 1
a compilation
of these surveys together with some additional information. This table 
must be considered a rough guide only to the current situation. However, it is indicative of the present state of our knowledge.

In Table 1 the first two columns give the name and the area searched for
each program. The third column is the limiting V magnitude of each survey. This
is a fairly crude estimate as some (eg Oppenheimer {\it et al.} 2001b) did not
survey in the V-band and it was necessary to transform their particular magnitude
to V so that a homogeneous set of statistics for each cluster could be derived.  Column 4
is an estimate of the distance probed in each 
survey for old white dwarfs. It is the distance out to which ancient white dwarfs could have 
been detected in the individual surveys. This limiting distance for detection
of old white dwarfs was set to $\rm 10(10^{([V_{lim} - 17]/5)})$ pc; 17
was taken as the $\rm M_V$ of a typical old white dwarf (Hansen 1998, 1999;
Saumon and Jacobsen 1999; Richer {\it et al.} 2000). Column 5 lists the number
of thick disk and spheroid white dwarfs that are expected out to the limiting distance
of the survey. This number was derived following the prescription of Reid
{\it et al.} (2001) and is dominated by the thick disk contribution. Thin disk white dwarfs are not

\begin{deluxetable}{c c c c c c c}
\tablenum{1}
\tablewidth{0pt}
\tablecaption{PREDICTED AND OBSERVED NUMBERS OF FAINT WHITE DWARFS}
\tablehead{
\colhead {Survey} & 
\colhead{Area (Deg$\rm^2$)} &
\colhead{$\rm V_{lim}$} &
\colhead {$\rm D_{max}$ (pc)} &
\colhead{No. Disk + Sph.} & 
\colhead{No. Dark Halo} & 
\colhead{No. Found}}
\startdata
\hline 
 Opp& 4165 & 19.8  & 36& 3.9  &  3.7 & 4 \\
\hline
Monet & 1378 & 20 & 40&  2.5 & 2.4 & $\rm >1$\\   
\hline
Ibata & 790 & 20 & 40& 1.4 & 1.4 & 2\\   
\hline
SDSS & 400 & 20.5 & 50&  1.4 & 3.5 & 2\\   
\hline
Jong & 2.5 & 24 &250& 1.1 & 3.6 & 3 \\   
\hline
CFHT & 15 &  25 & 400& 27 & 88 & ? \\
\hline
HDF & $\rm 1.4 \times 10^{-3}$ & 28 &1600 & 0.2 & 0.5 & 0? \\   
\hline

\enddata  

\noindent {\tablecomments {
Opp: Oppenheimer {\it et al.} 2001b, Science Online, March 22; 
Monet: Monet {\it et al.} 2000, AJ 120, 1541; 
Ibata: Ibata {\it et al.} 2000, ApJ 532, L41; 
SDSS: Harris {\it et al.} 2001, ApJ 549, L109; 
Jong: de Jong {\it et al.} 2000 astro-ph/0009058; 
CFHT: Stetson {\it et al.} in progress;
HDF: Ibata {\it et al.} 1999, ApJ 524, L95.}}
\end{deluxetable}

\noindent considered in this discussion as they would be eliminated by their low velocities. We are only concerned here with high velocity objects which could contaminate a true dark halo sample. Column 6 contains the number of observable white dwarfs expected
from the dark halo (Richer {\it et al.} 2000) assuming that the halo consists of 20\%
by mass of MACHOs which are old $\rm 0.5~M_{\odot}$ white dwarfs, half of which have hydrogen-rich outer atmospheres. The helium-rich ancient white dwarfs
would have cooled to very low luminosities by a Hubble Time and would be basically
unobservable. 
The seventh column lists the number of old white dwarfs actually found 
{\bf out to the limiting distance only} in each survey.

It is clear from examination of Table 1 that in all existing surveys thus
far, there is no obvious excess in the discovered number of old white
dwarfs over that expected from the thick disk and spheroid. Note, however,
that the CFHT program, currently acquiring its second epoch of data, should
be extremely important in this context. It is quite different from most of 
the 
other surveys as
it is much deeper ($\rm V_{lim} \sim 25$). One caveat must be kept in mind when perusing Table 1, however. In comparing, for example, columns 5 through 7 the assumption of 100\% recovery of all moving objects is built in. 
If, for example, a survey was only 50\% complete in recovering
moving objects, then the number in column 7 should be  increased
by a factor of 2. For some surveys, such as the SDSS, where proper motions
are not part of the selection criteria, the numbers in Table 1 are strictly correct. However,
until estimates for proper motion limits are accurately known, there remains no clear evidence of
a white dwarf dark matter halo signature in any of the current surveys. Of course, at the moment, it is also true that the surveys do not rule out
such a dark matter halo either.

\section{Temperature and Age Distribution of Dark Halo White Dwarf Candidates}

We can examine the totality of objects in the surveys of Table 1 from a different perspective. If they
are, in fact, local examples of the MACHOs, and thus important Galactic dark matter contributors, they should be quite old, likely
as old as the globular clusters. This would 
place them in the $\rm 12 - 14$ Gyr range if we
take the current spread quoted for globular cluster ages. From the location
of the white dwarfs in color-magnitude and color-color diagrams, temperatures
can be inferred with the use of theoretical models. From these temperatures,
ages can then be assigned using models of cooling white dwarfs after choosing
an appropriate mass. 

For the purposes of this exercise, I used the models of Hansen (1998, 1999)
and those discussed in Oppenheimer {\it et al.} (2001b) with white dwarf masses
of $\rm 0.5~M_{\odot}$, to estimate the temperature and cooling time of each object
found in the various surveys. Note that we are not limiting ourselves to objects within $\rm D_{max}$ here. The cumulative distribution of temperatures is shown in Figure
5 and the time for a $\rm 0.5~M_{\odot}$ white dwarf to cool to a given temperature is also indicated by the vertical lines. The 
immediate conclusion from the plot is that the sample 
in Table 1 is {\it not}
that of a halo sample, the objects are simply too young. It seems more in line with that of a thick disk component which dominates the predicted numbers
in Table 1. This result is, however, strongly model dependent. A different set of cooling
models or mass would provide quite a different result. For example, if I had chosen to assign a mass of $\rm 0.6~M_{\odot}$ to all the stars, the coolest object would have a cooling age of $\rm 11.7$ Gyr, almost $\rm 1$ Gyr older than a $\rm 0.5~M_{\odot}$
white dwarf. Further, the main sequence lifetimes of the white dwarf
precursors are not included in the cooling ages. If the masses of the precursors are in the range of $\rm 2~M_{\odot}$ this will add about 1 Gyr to these numbers. Nevertheless, the general conclusion to be drawn from both Table 1 and Figure 5 is that within the context of the present generation of white dwarf cooling models, truly ancient white dwarfs do not currently appear to have been discovered.

\begin{figure}
\psfig{figure=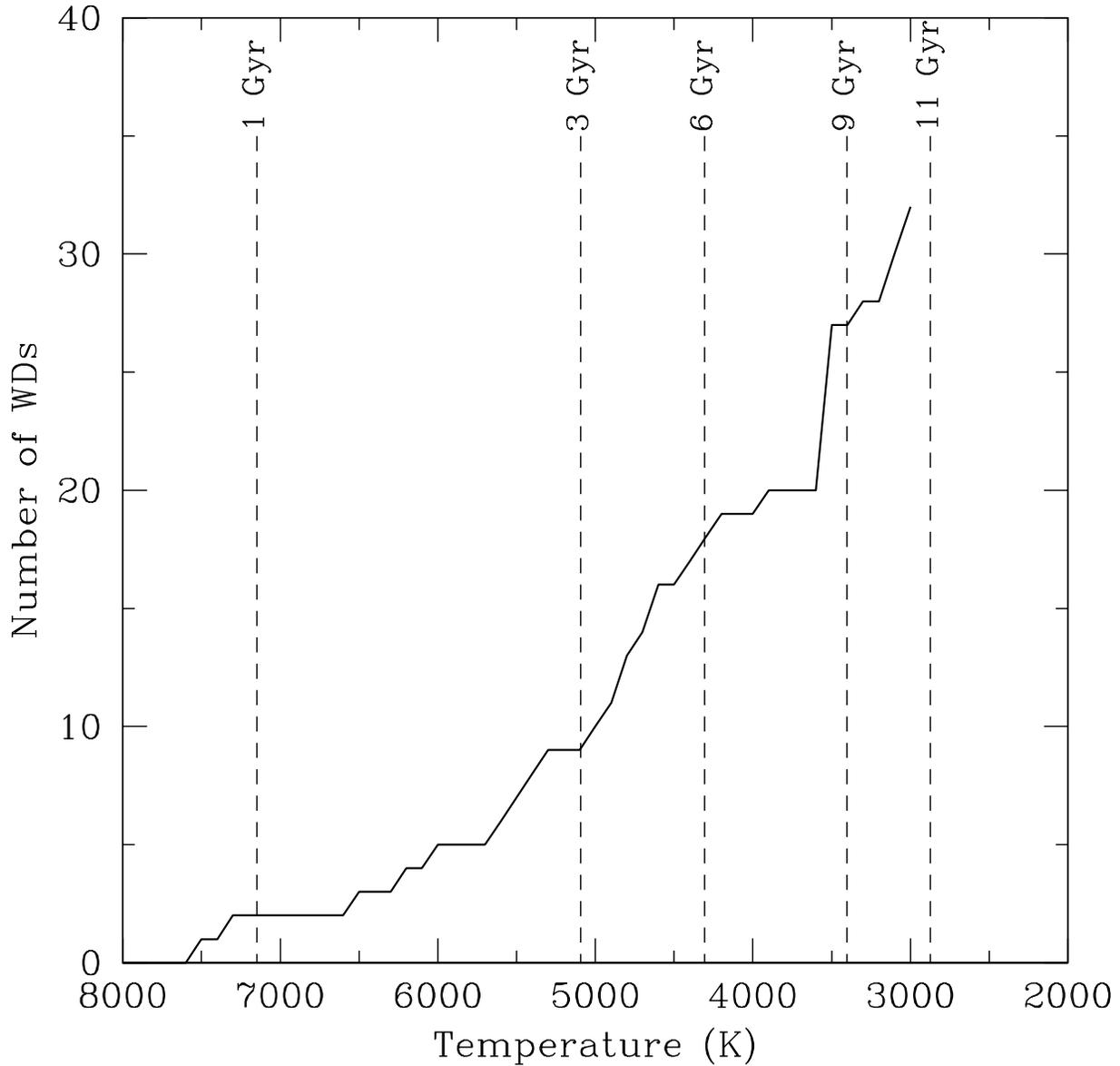,width=\hsize}
\caption{Cumulative distribution of the temperatures of all the white dwarfs with published
colors from the surveys listed in Table 1. Here I have selected not just
white dwarfs out to $\rm D_{max}$ for each survey, but the complete sample.
The time for a $\rm 0.5~M_{\odot}$ white dwarf to cool to
a given temperature is indicated. The age distribution seems to come
from a sample of objects that is significantly younger than the halo.}
\end{figure}

\section{Moving Objects in The Hubble Deep Field}

The last row in Table 1 deserves some special comments. Based on 2 epochs of
exposures taken in
the HDF separated by 2 years, Ibata {\it et al}. (1999) claimed to detect 2 extremely faint objects with significant
proper motion. One of these appeared to have colors consistent with that
expected from a very cool white dwarf. With lower statistical significance,
3 other objects also appeared to be moving, one of which possessed 
colors expected from an old white dwarf. The inferred 
space velocities for these objects
placed them squarely in the category of halo objects. This result was potentially very important as
the detection of even a few such objects in the HDF would be enough to account
for the entire MACHO contribution to the Galactic dark matter.
The second epoch data were, however, taken in the non-optimal F814W filter for another program (supernova search), and were quite noisy.

Based on this seemingly important result, a third epoch was secured in F606W,
fully 5 years after the original HDF data. These images were of significantly
higher quality. One of the objects suggested by Ibata {\it et al.} (1999) as possibly
moving was number 4-141 in the Williams {\it et al.} (1996) compilation. Using
a maximum likelihood technique, Ibata
{\it et al.} (1999) measured a motion of 35 mas/yr with an error of 8 mas/yr. With
the current superior data set we measure a proper motion of $\rm 10 \pm 4$ mas/yr,
a proper motion which is not statistically significant. None of the other 4 objects
was found to possess significant motion over the 5 year baseline. 

A full discussion of our reanalysis of moving objects in the HDF will appear
in Richer {\it et al.} 2002. 
 
\section{A Final Comment}

My personal view of the current landscape regarding the viability of old white dwarfs as a component of the Galactic dark matter is that while there are some
interesting hints, there is no compelling evidence in the current statistics either for against the scenario. Perhaps
a listing of the critical issues will be instructive.

\begin{enumerate}

\item {{\it Where are the lenses in the direction of the LMC and SMC located?}}
This is a fundamental issue $-$ perhaps the most important one. There are suggestions
that some, perhaps even most, of the lenses are in the Clouds themselves. This issue
may be
settled in a number of ways, some of which are possible with current technology.
One road to a solution is to observe along a different line of sight $-$ a microlensing
campaign in the direction of M31 or M33 would be extremely useful here. 
There are programs currently
underway looking at M31 (e.g. Uglesich {\it et al.} 1999) but no comprehensive results are
available at this time. Some future satellite missions may be capable of detecting
parallaxes (and masses) for the lenses (Gould and Salim 1999) and this would be critical data for
establishing their location. If the lenses are
in the Clouds and are low mass main sequence stars, then it may eventually be
possible to observe them directly.

\vspace{0.2in}

\item {{\it Is there a population of cool white dwarfs with halo kinematics that are present in numbers which exceed those expected from the thick disk and spheroid?} As I have discussed, based solely on number count arguments, the current searches do not as yet
provide a conclusive answer either in the affirmative or negative.
Deep ($\rm V \sim 25$) proper motion and color selected surveys covering 10 degrees or more of sky are required
to be definitive. There are several such searches underway and the results
are eagerly anticipated.}

\vspace{0.2in}

\item {{\it Do we have the last word on the models of old white dwarfs?} There are 2 aspects of the present theory of cooling white dwarfs which I find disturbing. First, there are enormous problems in fitting the spectra of these cool white dwarfs with 
the current generation of models. The situation is
so dire that Bergeron (2001), one of the premier practitioners of this art, was
required to insert a fictitious opacity source into his models to help in the
fits. He then went on to state ``It now remains to be seen whether this new opacity source, even when included with the approximate treatment introduced in this paper, can help resolve the mystery surrounding several ultracool white dwarfs whose energy distributions have yet failed to be successfully explained in terms of hydrogen or mixed hydrogen/helium compositions." The second concern
is simply that no truly ultracool white dwarfs have as yet been discovered, or,
perhaps better put, no white dwarfs are known whose colors and/or spectra suggest that the object's age is in the range of $\rm 12 - 14$ Gyrs. This may be the
same problem as the first one mentioned here, that is that the models are still
somehow inadequate, or it may be suggesting that such objects are extremely rare
or just do not exist. It is within this context that extremely deep observations of globular
cluster white dwarfs will be important.}

\end{enumerate}

\begin{acknowledgements}

The author would like to thank the Space Telescope Science Institute for
organizing a meeting that was all that a scientific should be; exciting new
results, participants with deep and wide interests, an excellent venue. In
particular, deepest gratitude is due to Mario Livio who organized the meeting
and provided one of the best conference summaries that this author has ever
experienced.

The research of HBR is supported by grants from the Natural Sciences and Engineering Research Council of Canada.

\end{acknowledgements}

\end{document}